\documentclass[11pt]{article}%
\usepackage{graphicx}%
\usepackage{hyperref}
\usepackage{amsmath}
\usepackage{amscd,amssymb}
\def\cR{{\mathcal R}}

\newcommand{\beq}{\begin{eqnarray}}
\newcommand{\eeq}{\end{eqnarray}}
\numberwithin{equation}{section}

\begin{document}
\hfill{SISSA/27/2014/FISI}
\vspace{0.5in}

\begin{center}
{\large\bf Generalized $q$-deformed Correlation Functions as Spectral Functions of Hyperbolic Geometry}
\end{center}

\vspace{0.1in}

\begin{center}
{\large L. Bonora $^{(a)}$ \footnote{bonora@sissa.it},
A. A. Bytsenko $^{(b)}$ \footnote{aabyts@gmail.com},
and M. E. X. Guimar\~aes $^{(c)}$ \footnote{emilia@if.uff.br}

\vspace{6mm} $^{(a)}$ {\it International School for Advanced Studies (SISSA/ISAS)
Via Bonomea 265, 34136 Trieste and INFN, Sezione di Trieste, Italy}

\vspace{0.2cm} $^{(b)}$ {\it Departamento de F\'{\i}sica,
Universidade Estadual de Londrina, Caixa Postal 6001,
Londrina-Paran\'a, Brazil}

\vspace{0.2cm} $^{(c)}$ {\it Instituto de F\'{\i}sica,
Universidade Federal Fluminense,  Av. Gal. Milton Tavares de
Souza, s/n CEP 24210-346, Niter\'oi-RJ, Brazil}}
\end{center}

\vspace{0.1in}

\begin{abstract}
We analyse the role of vertex operator algebra and 2d amplitudes from the point of view of the representation theory of infinite dimensional Lie algebras, MacMahon and Ruelle functions.
A $p$-dimensional MacMahon function is the generating function of $p$-dimensional partitions of integers.
These functions can be represented as amplitudes of a two-dimensional $c=1$ CFT. In this paper we show that $p$-dimensional MacMahon functions can be rewritten in terms of Ruelle spectral functions, whose spectrum is encoded in the Patterson-Selberg function of three dimensional hyperbolic geometry.
\end{abstract}

\vspace{1.0in}
\begin{flushleft}

\end{flushleft}

\newpage

\section{Introduction}

This paper, whose main focus is the relation between MacMahon and Ruelle functions, is motivated by the steady, if not growing, interest in the application of symmetric functions, in particular of the two dimensional MacMahon function and its higher dimensional generalizations to physical systems. This occurs in many areas of statistical physics \cite{Okunkov,Kenyon2003,Kenyon2006} and topological string theory \cite{Ghostal,Witten,Saidi},
BPS black holes, model of branes wrapping collapsed cycles in Calabi-Yau orbifolds, and quiver gauge theories \cite{Douglas,Benhaddou,Saraikin}.
Generalized MacMahon functions are used, in particular, in computation of amplitudes
of A-model topological string \cite{Aganagis,Iqbal2003,Iqbal2007,Drissi}, more specifically to the so-called topological vertex.

A $p$-dimensional MacMahon function is the generating function of $p$-dimensional partition of integers, which is the number of different ways in which we can split an integer using distinct
p-dimensional arrays of other nonnegative integers. As we shall see these functions can be represented as amplitudes of a two-dimensional CFT. It is not clear whether the dimension $p$
of the partition can be assigned any physical meaning. But we will show in this paper that $p$-dimensional MacMahon functions can be rewritten in terms of Ruelle spectral functions, whose spectrum is encoded in the Patterson-Selberg function of three dimensional hyperbolic geometry. This may lead to an interpretation of the above results in terms of ADS/CFT correspondence, an attractive possibility which we leave for future investigation.

There is also another side of our work we would like to recall.
We have remarked elsewhere
the important connection between quantum generating functions in physics and
formal power series associated with dimensions of chains and homologies of suitable infinite dimensional Lie algebras. MacMahon and symmetric functions play an important role in the homological aspects of this connection; its application to partition functions of minimal three-dimensional gravities in the space-time
asymptotic to AdS$_3$, which also describe the three-dimensional Euclidean black holes, pure
supergravity, elliptic genera and associated $q$-series were studied in \cite{BB,BBE}.
On the other hand special applications of symmetric functions appear in the representation theory of infinite dimensional
Lie algebras  \cite{Kac,Awata,BB,BBE}.
The utility of symmetric function techniques can be demonstrated in providing concrete realizations of the (quantum) affine algebra, for instance in calculating the trace of products of currents of this algebra. These functions are
respectively the appropriate character ${\rm ch}_{\mathbb R}$ of the basic representations of the ${\mathfrak s}{\mathfrak l}(\infty) $ and the affine algebra $\widehat{{\mathfrak s}{\mathfrak l}(\infty )}$
at large central charge $c$ \cite{Frenkel}. Note that all simple (twisted and untwisted) Kac-Moody algebras can be embedded in the ${\mathfrak s}{\mathfrak l}(\infty)$ algebra of infinite matrices with a finite number of non-zero entries, which has a realization in terms of generators of a Clifford algebra.
It has been observed that in the limit $c\rightarrow \infty$ the basic representation of
$\widehat{{\mathfrak s}{\mathfrak l}(\infty )}$ is related to the partition function of a three
dimensional field theory \cite{Douglas,Heckman}.

Needless to say, although these links are suggestive, the general panorama looks still inarticulate, and more models and examples are needed to accommodate them into a precise scheme.
One of the purposes of the present paper is to better understand the role of vertex algebra and 2d amplitudes from the point of view of the representation theory of infinite dimensional Lie algebras, MacMahon and Ruelle functions. In this regard a particular  importance  the correspondence between Ruelle spectral functions and Poincar\'{e} $q$-series associated
with conformal structure in two dimensions.

The organization of the paper is as follows. In Sect. 2 we introduce the algebra of $q$-deformed vertex operators (of the $c=1$ 2d-conformal model) and consider their generalizations and properties essential for the next sections.
In Sect. 3 we reformulate the generalized MacMahon functions in terms of the Ruelle
spectral functions of hyperbolic geometry. We also broach and briefly discuss
the topic of higher-dimensional partitions, as such, originally introduced by MacMahon.
We analyze correlation functions of vertex operators, the MacMahon's conjecture
(Eq. (\ref{mu}) and their possible interpretation as $p$-dimensional partition functions.

In Sect. 4 we consider multipartite (vector valued) generating functions and utilize well-known formulas
for Bell polynomials. We derive the infinite hierarchy of $q$-deformed vertex operators and factorized partition
functions and represent them by means of spectral functions.

Finally in Sect. 5 we conclude with summarize the main results accompanied with discussions and suggestions.

In the Appendix we give a few formulas involving Ruelle and Patterson-Selberg spectral functions of hyperbolic three-geometry.

\section{Algebra of vertex operators}

In this section we introduce the notation and quote some earlier results we need. To this end
we follow mostly \cite{Iqbal2007} and, in particular, the subsequent elaboration by \cite{Drissi2008}.
Let us consider the hierarchy of generalized local $q$-deformed vertex operators $\Gamma_{\pm}^{(p)}(z, q)$\,
$(p>1)$
\begin{equation}
\Gamma_{\pm}^{(p)}(z, q) = \exp \left(\sum_{n=1}^\infty \frac{\mp i}{n}\frac{z^{\mp n}}{(1-q^n)^{p-1}} J_{\pm n}\right)\,.
\end{equation}
$J_n$ are the modes of a standard holomorphic $U(1)$ Kac-Moody $J(z)$ with Laurent expansion is $J(z)= \sum_{n\in {\mathbb Z}}z^{-n-1}J_n,\,
J_n = \oint (z^n/2\pi i)J(z)dz$. The Heisenberg algebra is $[J_n, J_m]= n\delta_{n+m, 0},\, J_n^\dagger = J_{-n}$ and $J_n\vert0\rangle=0$ for $n\geq 1$. The commuting mode $J_0$ is disregarded.

One can use the identities
$
z^{\pm n}/(1-q^n)^{p-1}  =  \sum_{j=1}^{p-1}z^nq^{\pm n}/(1-q^n)^j +z^{\pm n}
$
to obtain the recursive relations
\begin{equation}
\Gamma_{-}^{(p)}(z, q) = \Gamma_{-}^{(1)}(z)\prod_{j=2}^{p-1}\Gamma_{-}^{(j)}(qz, q)\,,
\,\,\,\,\,\,\,\,\,
\Gamma_{+}^{(p)}(z, q)  =  \Gamma_{+}^{(1)}(z)\prod_{j=2}^{p-1}\Gamma_{+}^{(j)}(z/q, q)\,.
\label{gv}
\end{equation}

The local operators $\Gamma _{\pm }^{(1)}(z):= \Gamma_{\pm}(z)$ act on the Hilbert space
states of the $c=1$ 2d-conformal field theory and exhibit properties inherited from the algebra of the $J_{\pm n}$.
They obey the following algebra
\begin{eqnarray}
\Gamma _{\pm }(x)\Gamma _{\pm }(y) & = &\Gamma _{\pm }(y)\Gamma _{\pm}(x),
\,\,\,\,\,\,\,\,\,\,\,\, x,y\in {\mathbb C},
\nonumber \\
\Gamma _{+}(x)\Gamma _{-}(y) & = & (1-y/x)^{-1}\Gamma_{-}(y) \Gamma _{+}(x)\,.
\end{eqnarray}

It is interesting to note that introducing $L_0= \sum_{n>0}^\infty J_{-n}J_n$, we get
\begin{eqnarray}
[L_0, \Gamma_{\pm}^{(p)}(z, q)]= z\frac d{dz} \Gamma_{\pm}^{(p)}(z, q).
\end{eqnarray}
Thus $\Gamma_{\pm}^{(p)}(z, q)$ are `weight 0 primaries'. It follows, in particular, that
the operator $q^{L_{0}}$ acts on $\Gamma _{\pm}(z,q)$ as follows:
$q^{L_{0}}\Gamma _{\pm }(z,q)q^{-L_{0}}=\Gamma_{\pm }(qz,q)$.

Due to the properties $J_{n}|0\rangle =0$, $\langle 0|J_{-n}=0$ for $n>0$,
the operators $\Gamma _{\pm }(z)$ act on the vacuum as the identity operator:
$
\Gamma _{+}(z)|0\rangle = |0\rangle, \langle 0|\Gamma _{-}(z)= \langle 0|
$
it follows that
\begin{eqnarray}
 \langle 0|\Gamma _{+}(z)\Gamma _{-}(w)|0\rangle= \frac 1{1-\frac wz}.
\end{eqnarray}

As noted in \cite{Drissi2008}, $\Gamma _{-}(z)$ contains all the monomials in $J_{-n_{j}}$,
$J_{-\mathbf{n}}^{\lambda }\equiv \prod_{j\geq 1}\left( J_{-n_{j}}\right)^{\lambda _{j}},$
where $\lambda =\left( \lambda _{1},\lambda _{2},\ldots \right)$ is a 2d partitions; the state
$\Gamma _{-}(z)|0\rangle$ is reducible and is given by a sum over all possible 2d partitions $\lambda $.
In the case $z=1$,\,\,
$
\Gamma _{-}(1)|0\rangle =\sum_{(\text{{\small 2d partitions} }\lambda)}|\lambda \rangle.
$
A similar relation is valid for $\langle 0|\Gamma _{+}(1)$.

With the mathematical tools just introduced we can proceed to higher dimensional generalizations. For example, one can rewrite $\Gamma_{\pm}^{(2)}(z, q)$ as follows:
\begin{eqnarray}
\!\!\!\!\!
\Gamma_{-}^{(2)}(z, q) & = & \prod_{t=-\infty }^{-1}\Gamma
_{-}(1)\left( z\right) q^{L_{0}} =\prod_{k=0}^{\infty }\Gamma
_{-}(1)\left( q^{k}z\right)
=  \exp \left( \sum_{n\geq 1}\frac{i}{n}\frac{z^{n}
}{\left( 1-q^{n}\right) }J_{-n}\right),
\label{G+}
\\
\!\!\!\!\!
\Gamma_{+}^{(2)}(z, q) & = & \prod_{t=0}^{\infty
}q^{L_{0}}\Gamma _{+}(1)\left(z \right) =\prod_{k=0}^{\infty
}\Gamma _{+}(1) \left( q^{-k}z\right)
= \exp \left( -\sum_{n\geq 1}\frac{i}{n}\frac{%
z^{-n}}{\left( 1-q^{n}\right) }J_{n}\right).
\label{G-}
\end{eqnarray}
The products $\prod_{t=-\infty}^{-1}(-\!\!-)$ in these equations are taken over diagonal slices of the 3d partitions. They are reminiscent of the transfer matrix method where a 3d partition is thought of as an amplitude
between the slice at $t=-\infty$ ({\it in-state}) and the slice at $t=\infty$
({\it out-state}). The algebra of these vertex operators takes the following form
\begin{equation}
q^{L_{0}}\Gamma_{\pm }^{(2)}( z, q) q^{-L_{0}}= \Gamma_{\pm }^{(2)}( qz, q),\,\,\,\,\,\,\,\,
\Gamma_{\pm }^{(2)}( z, q)\Gamma_{\pm }^{(2)}(w, q) =\Gamma_{\pm }^{(2)}(w, q)\Gamma_{\pm }^{(2)}( z, q).
\end{equation}

We can replicate recursively this construction. This leads us to the following hierarchy of composite vertex operators
\begin{equation}
\Gamma _{-}^{\left( n+1\right) }\left( z, q\right) =
\prod_{t_{n}=1}^{\infty }\cdots  \prod_{t_{2}=1}^{
\infty } \prod_{t_{1}=1}^{\infty }\left(\Gamma _{-}\left( z\right)
q^{L_{0}}\right)\cdot q^{L_{0}}\cdots q^{L_{0}}.\label{Gammazq}
\end{equation}
A similar expression can be written down for $\Gamma _{+}^{\left( n+1\right) }\left( z, q\right) $. For $n=0,$ we have just $\Gamma _{-}\left( z\right) $.
It is not difficult to check that the explicit expression of the vertex operators $\Gamma _{-}^{\left( p\right)
}\left( z, q\right) $  ($p\geq 1$) acting on the vacuum is given by
\begin{eqnarray}
\!\!\!\!\!
\Gamma _{-}^{\left( p\right) }\left( z, q\right) \left\vert 0\right\rangle
=  \exp \left( \sum_{n=1}^{\infty }\frac{iz^{n}J_{-n}}{n\left( 1-q^{n}\right)
^{p-1}}\right) \left\vert 0\right\rangle
=\Gamma _{-}\left( z\right)
\prod_{k=2}^{p-1}\Gamma _{-}^{\left( k\right) }\left( qz, q\right)
\left\vert 0\right\rangle.
\nonumber
\end{eqnarray}

\section{MacMahon, partitions, and Ruelle spectral functions revised}

In this section, using the formulas in Appendix, we transcribe the MacMahon partition functions in terms of spectral functions of hyperbolic geometry.

The 1d- and 2d-MacMahon function can be interpreted as the two-point correlation of the vertex
operators $\Gamma _{+}\left( 1\right)$ and $\Gamma _{-}\left(q\right)$
\begin{eqnarray}
{Z}_{1d} & = &\left\langle 0|\Gamma _{+}\left( 1\right) \Gamma _{-}\left( q\right)
|0\right\rangle =
\left\langle 0|\text{ }\Gamma _{+}\left( 1\right) q^{L_{0}}\Gamma
_{-}\left( 1\right) |0\right\rangle = (1-q)^{-1},
\\
{Z}_{2d} & = & \langle 0|\Gamma _{+}\left( 1\right) q^{L_{0}}
\prod_{k\geq 1}\Gamma _{-}( 1) q^{L_{0}}|0\rangle
=
\langle 0|\Gamma _{+}(1)\prod_{k\geq 1}\Gamma_{-}(q^{k})\vert 0\rangle
\nonumber \\
& = & \prod_{k\geq 1}\left( 1-q^{k}\right) ^{-1} \stackrel{{\rm by\, Eq.\, (\ref{R1})}}{=\!=\!=\!=\!=\!=}
[{\mathcal R}(s= 1-i\varrho(\tau))]^{-1}.
\end{eqnarray}
In the previous section we have introduced  hierarchy of level $p$
vertex operators $\Gamma_{\pm}^{(p)}$. In perfect analogy with ${Z}_{1d}$ and ${Z}_{2d}$ we can introduce and compute ${Z}_{3d}$:
\begin{eqnarray}
{Z}_{3d} & = & \left\langle 0|\left( \prod_{t=0}^{\infty }q^{L_{0}}\Gamma
_{+}(1)\right) q^{L_{0}}\left( \prod_{t=-\infty }^{-1}\Gamma
_{-}(1)q^{L_{0}}\right) |0\right\rangle
\nonumber \\
& = & \left\langle 0|\text{ }\prod_{t=0}^{\infty }\Gamma _{+}^{(1)}\left(
q^{-t-\frac{1}{2}}\right) \prod_{\ell =1}^{\infty }\Gamma _{-}^{(1)}\left(
q^{\ell-\frac{1}{2}}\right) \text{ }|0\right\rangle
\nonumber \\
& = & \prod_{\ell=0}^{\infty }\prod_{j=1}^{\infty }\left[1-q^{j+\ell}\right]^{-1}
\stackrel{k:=j+\ell}{=\!=\!=\!=}  \prod_{k=1}^{\infty }\prod_{j=1}^{k}
\left[1-q^{k}\right]^{-1}
=\prod_{k=1}^{\infty }\left[1-q^{k}\right]^{-k}\!\!.
\label{Z1}
\end{eqnarray}
When obtaining the second line in Eq. (\ref{Z1}) we split $q^{L_{0}}$ as $q^{L_{0}/2}q^{L_{0}/2}$
and commute each of the operators $q^{L_{0}/2}$ to the left and the other to the right.
The last product in Eq. (\ref{Z1}) is precisely the usual form of the 3d-MacMahon function,
which, again, can be rewrite in terms of the spectral Ruelle functions:
\begin{equation}
\prod_{k=1}^{\infty }\left[1-q^{k}\right]^{-k} \stackrel{{\rm by\, Eq.\, (\ref{R3})}}{=\!=\!=\!=\!=\!=}
\prod_{n=1}^{\infty}[{\mathcal R}(s= n(1-i\varrho(\tau)))]^{-1}\,.
\end{equation}

{\bf $p$-dimensional partition functions.} The structure of the $p$-dimensional partition function ${Z}_{pd}$
can be analyzed in terms of the vertex operators $\Gamma _{\pm }^{(p)}(z, q)$ introduced before.
As we have seen above, (\ref{Gammazq}), the latter  can be interpreted as the level $p$
generalization of $\Gamma _{-} (z)$, they obey the relations
$
\Gamma _{-}^{\left( p\right) }\left( z, q\right) =q^{L_{0}}\Gamma _{-}^{\left(
p\right) }\left( 1, q\right) q^{-L_{0}},\,p\geq 0.
$
The $p$-dimensional partition functions $Z_{pd}$ can be defined as \cite{Drissi2008}
\begin{equation}
{Z}_{pd}=\left\langle 0|\Gamma _{+}\left( 1\right) \Gamma _{-}^{\left(
p\right)}(z, q) |0\right\rangle
\equiv
\left\langle 0|\Gamma _{+}\left( 1\right) q^{L_{0}}\Gamma
_{-}^{\left( p\right) }\left( 1, q\right) |0\right\rangle, \,\,\,\,\,\,\,
p\geq 0.
\label{pd}
\end{equation}
We can rewrite these partition functions in terms of Ruelle spectral function.
Indeed, commuting $\Gamma _{-}^{\left( p\right) }\left( z, q\right) $ to the left of
$\Gamma _{+}\left( 1, q\right) $ for $p\geq 2$, one can get by induction (see for detail
\cite{Drissi2008}) that
\begin{eqnarray}
{Z}_{pd} & = & \prod_{k=1}^{\infty}\left[1-q^{k}\right]^{-C(k, p)}
\stackrel{{\rm by\,\, Eq. (\ref{G1})}}{=\!=\!=\!=\!=\!=}
\prod_{n=1}^{\infty}\left[\frac{{\mathcal R}(s= n(1-i\varrho(\tau)))}
{{\mathcal R}(s= (n+1)(1-i\varrho(\tau)))}\right]^{-C(n, p)}
\label{Fp},
\\
C(n, p)\!\!\! & = & \frac{(n+p-3)!}{(n-1)!( p-2)!}.
\label{Cp}
\end{eqnarray}

{\bf Plane partitions.} So far we have called $Z_{pd}$ a p-dimensional partition function without any comment. Here we would like to motivate this term.
It comes from the fact that correlation functions of the corresponding vertex operators admit a presentation that can be associated with higher-dimensional partitions. Recall that higher-dimensional partitions of $n$ is an array of numbers whose sum is $n$:
\begin{equation}
n = \sum_{j_1, \ldots, j_r\geq 0} n_{j_1j_2\ldots j_r},\,\,\,\,\,
{\rm where}\,\,\,\,\, n_{j_1j_2\ldots j_r}\geq n_{k_1k_2\ldots k_r}
\end{equation}
whenever $j_1\geq k_1, j_2\geq k_2, \ldots, j_r\geq k_r$, and all $n_{j_1j_2\ldots j_r}$  nonnegative integers.
Let us introduce also plane partitions: these are two-dimensional arrays of nonnegative integers subject to a nonincreasing condition along rows and columns. It is worth recalling that Young tableaux with strict decrease along columns are essentially equivalent to plane partitions. They were originally used by Alfred Young in his work on invariant theory. Young tableaux have played an important role in the representation theory of the symmetric group; they also occur in algebraic geometry and in many combinatorial problems.

Let us denote $\pi_r(n_1, n_2, \ldots, n_k; q)$ the generating function for plane
partitions with at most $r$ columns, at most $k$ rows, and with $n_i$ the first entry in the $i$th row.
The functions $\pi_r(n_1, n_2, \ldots, n_k; q)$ are completely determined by
the following recurrence and initial condition
\begin{eqnarray}
\pi_{r+1}(n_1, n_2, \ldots, n_k; q) & = & q^{\sum_{j=1}^k(n_j)}
\sum_{m_k=0}^{n_k} \sum_{m_{k-1}= m_k}^{n_k-1}\cdots \sum_{m_1= m_2}^{n_1}
\pi_r(m_1, \ldots , m_k; q),
\label{p1}
\\
\pi_1(n_1, n_2, \ldots, n_k; q) & = &  q^{\sum_{j=1}^k(n_j)}.
\label{p2}
\end{eqnarray}
$\pi_{r+1}(n_1, n_2, \ldots, n_k; q)$ can be represent as a determinant (see for detail \cite{Andrews}):
\begin{equation}
\pi_{r}(n_1, n_2, \ldots, n_k; q)  =  q^{\sum_{j=1}^k(n_j)}
{\rm det}\left[q^{(i-j)(i-j-1)/2} \begin{pmatrix}
n_j+r-1\cr r-i+j-1\end{pmatrix}  \right]_{1\leq i,\,j\leq k}\!\!\!\!\!\!\!\!.
\label{det}
\end{equation}
Define the number of plane partitions of $m$, $p_{k,r}(m, {n})$, with at most $r$ columns,
at most $k$ rows, and with each entry $\leq {n}$, and let
$
\pi_{k,r}({n}; q) := \sum_{m=0}^\infty p_{k,r}(m, {n})q^m.
$
Then one can observe that
\begin{eqnarray}
\pi_{k,r}({n}; q) & \stackrel{{\rm by\, Eq.\,(\ref{p1})}}{=\!=\!=\!=\!=\!=\!=}&
\sum_{n_k\leq \cdots \leq n_1\leq n} \pi_{r}(n_1, n_2, \ldots, n_k; q)
= q^{-kn}\pi_{r+1}(n_1, n_2, \ldots, n_k; q)
\nonumber \\
& \stackrel{{\rm by\, Eq.\,(\ref{det})}}{=\!=\!=\!=\!=\!=\!=}&
{\rm det} \left[q^{(i-j)(i-j-1)/2} \begin{pmatrix}
n+r\cr r-i+j\end{pmatrix}  \right]_{1\leq i,\,j\leq k}\!\!\!\!\!\!\!\!.
\end{eqnarray}
As a result MacMahon's formulas for the
generating function of $k$-rowed plane partitions $\pi_{k,\infty}(\infty; q)$ follows \cite{Andrews}
\begin{eqnarray}
\sum_{m=0}^\infty p_{k, \infty}(m, \infty)q^m & = & \prod_{n=1}^\infty (1-q^n)^{-{\rm min}(k,n)},
\\
\sum_{m=0}^\infty p_{\infty, \infty}(m, \infty)q^m & = & \prod_{n=1}^\infty (1-q^n)^{-n} = Z_{3d}.
\end{eqnarray}
Let $\mu_k(j)$ be the number of $k$-dimensional partitions of $j$, then due to the MacMahon's
conjecture
\begin{equation}
\sum_{j=0}^\infty \mu_k(j) q^j = \prod_{n=1}^\infty (1-q^n)^{-C(n,k)},\,\,\,\,\,\,\,\,\,
C(n,k) = \frac{(n+k-2)!}{(n-1)!}\,.
\label{mu}
\end{equation}
MacMahon eventually came to doubt the truth of (\ref{mu}) in general; in fact,
its falsehood in general was established in the late 1960's, \cite{Andrews} . However
this conjecture is certainly true for $k = 1$ and 2. It is remarkable that
\begin{eqnarray}
\sum_{j=0}^\infty \mu_1(j)q^j & = & \prod_{n=1}^\infty(1-q^n)^{-1}= Z_{2d}\,,
\\
\sum_{j=0}^\infty \mu_2(j)q^j & = & \prod_{n=1}^\infty(1-q^n)^{-n}= Z_{3d}\,.
\end{eqnarray}
Comparing (\ref{mu}) and (\ref{Fp}), (\ref{Cp}) we get the relations
$C(n, k= 1) = C(n, p=2)$, \, $C(n, k=2)= C(n, p=3)$. The MacMahon`s conjecture for the case $k> 2$
can be corrected by using of the comparison between $C(n,k)$ and the power $C(n,p)$ in
the $q$-expansion of $p$-dimensional partition function $Z_{pd}$.

\section{Multipartite generating functions and infinite hierarchy of $q$-deformed vertex operators}

{\bf Multipartite generating functions.}
Let consider, for any ordered $\ell$-tuple of nonnegative integers not all zeros, $(k_1, k_2, \ldots ,k_\ell)={\bf k}$ (referred to as "$\ell$-partite" or {\it multipartite} numbers), the
(multi)partitions, i.e. distinct representations of $(k_1, k_2, \ldots ,k_\ell)$ as sums of multipartite numbers. Let us call
${\mathcal C}_-^{(u,\ell)}({\bf k}) = {\mathcal C}_-^{(\ell)}(u;k_1, k_2 , \cdots , k_\ell)$
the number of such multipartitions, and introduce in addition the symbol
${\mathcal C}_+^{(u,\ell)} ({\bf k})= {\mathcal C}_+^{(\ell)}(u;k_1, k_2 , \cdots , k_\ell)$.
Their generating functions are defined by
\begin{eqnarray}
{\mathcal F}(u) &: = & \prod_{{\bf k}\geq 0} \left( 1- ux_1^{k_1}x_2^{k_2}\cdots
x_\ell^{k_\ell}\right)^{-1} = \sum_{{\bf k}\geq 0}{\mathcal C}_-^{(u,\ell)}({\bf k})
x_1^{k_1}x_2^{k_2}\cdots x_\ell^{k_\ell}\,,
\label{PF1}
\\
{\mathcal G }(u) &: = & \prod_{{\bf k}\geq 0} \left( 1 + ux_1^{k_1}x_2^{k_2}\cdots
x_\ell^{k_\ell}\right) = \sum_{{\bf k}\geq 0}{\mathcal C}_+^{(u,\ell)}({\bf k})
x_1^{k_1}x_2^{k_2}\cdots x_\ell^{n_\ell}\,.
\label{PF2}
\end{eqnarray}
Therefore,
\begin{eqnarray}
\!\!\!\!\!\!\!\!\!\!\!\!\!\!\!\!
{\rm log}\, {\mathcal F}(u) & = & - \sum_{{\bf k}\geq 0} {\rm log}
\left(1- ux_1^{k_1}x_2^{k_2}\cdots x_\ell^{k_\ell}\right)
=  \sum_{{\bf k}\geq 0} \sum_{m=1}^\infty \frac{u^m}{m}
x_1^{mk_1}x_2^{mk_2}\cdots x_\ell^{mk_\ell}
\nonumber \\
& = &
\sum_{m =1}^\infty \frac{u^m}{m}
(1-x_1^m)^{-1}(1-x_2^m)^{-1} \cdots (1-x_\ell^m)^{-1}
=  \sum_{m=1}^\infty \frac{u^m}{m}
\prod_{j= 1}^r (1-x_j^m)^{-1},
\\
\!\!\!\!\!\!\!\!\!\!\!\!\!\!\!\!
{\rm log}\,{\mathcal G}(-u) & = & {\rm log}\,{\mathcal F}(u)\,.
\end{eqnarray}
Finally,
\begin{eqnarray}
\!\!\!\!\!\!\!\!\!\!\!\!\!
{\mathcal F}(u) & = & \sum_{{\bf k}\geq 0}{\mathcal C}_-^{(u,\ell)}({\bf k})
x_1^{k_1}x_2^{k_2}\cdots x_\ell^{k_\ell}
= \exp\left( \sum_{m=1}^\infty \frac{u^m}{m}
\prod_{j=1}^r (1-x_j^m)^{-1}\right),
\\
\!\!\!\!\!\!\!\!\!\!\!\!\!
{\mathcal G }(u) & = & \sum_{{\bf k}\geq 0}{\mathcal C}_+^{(u,\ell)}({\bf k})
x_1^{k_1}x_2^{k_2}\cdots x_\ell^{n_\ell}
= \exp\left( \sum_{m=1}^\infty \frac{(-u)^m}{m}
\prod_{j=1}^r (1-x_j^m)^{-1}\right).
\end{eqnarray}
It is known that the Bell polynomials are very useful in many problems in combinatorics. We would like to note their
application in multipartite partition problem \cite{Andrews}. The Bell polynomials technique can be used for
the calculation ${\mathcal C}_-^{(\ell)}({\bf k})$ and ${\mathcal C}_+^{(\ell)}({\bf k})$.
Let
\begin{eqnarray}
&& {\mathcal F}(u) := 1 + \sum_{j=1}^\infty {\mathcal P}_j(x_1,x_2, \ldots, x_\ell)u^j,
\,\,\,\,\,\,\,\,\,\,
{\mathcal P}_j  =  1+ \sum_{{\bf k}> 0}P({\bf k}; j)x_1^{n_1}\cdots x_\ell^{n_\ell},
\label{Fu}
\\
&& {\mathcal G}(u)  :=  1 + \sum_{j=1}^\infty {\mathcal Q}_j(x_1,x_2, \ldots, x_\ell)u^j,
\,\,\,\,\,\,\,\,\,\,
{\mathcal Q}_j  =  1+ \sum_{{\bf k}> 0}Q({\bf k}; j)x_1^{n_1}\cdots x_\ell^{n_\ell}.
\label{Gu}
\end{eqnarray}
Useful expressions for the recurrence relation of the Bell polynomial $Y_{n}(g_1, g_2, \ldots , g_{n})$
and generating function ${\mathcal B}(u)$ have the forms \cite{Andrews}:
\begin{eqnarray}
&& Y_{n+1}(g_1, g_2, \ldots , g_{n+1}) = \sum_{k=0}^n  \begin{pmatrix} n\cr k\end{pmatrix}
Y_{n-k}(g_1, g_2, \ldots , g_{n-k})g_{k+1},
\label{B2}
\\
&& {\mathcal B}(u) = \sum_{n=0}^\infty \frac{Y_nu^n}{n!} \,\,\,\,\,\, \Longrightarrow \,\,\,\,\,\,
{\rm log}\,{\mathcal B}(u)= \sum_{n=1}^\infty \frac{g_nu^n}{n!}\,.
\label{B1}
\end{eqnarray}
To verify the second formula in (\ref{B1}) we need to differentiate with respect to $u$ and observe
that a comparison of the coefficients of $u^n$ in the resulting equation produces an
identity equivalent to (\ref{B2}). From Eq. (\ref{B2}) one can obtain the following explicit
formula for the Bell polynomials (it is known as Faa di Bruno's formula)
\begin{equation}
Y_{n}(g_1, g_2, \ldots , g_{n}) = \sum_{{\bf k}\,\vdash\, n}\frac{n!}{k_1!\cdots k_n!}
\prod_{j=1}^n\left(\frac{g_j}{j!}\right)^{k_j}\!\!.
\end{equation}
Let $\beta_r(m) := \prod_{j=1}^r(1-x_j^m)^{-1}$; the following result holds
(see for detail \cite{Andrews}):
\begin{eqnarray}
{\mathcal P}_j  & = & \frac{1}{j!}Y_j \left( 0!\beta_r(1),\,\, 1!\beta_r(2)\,\,,
\ldots , \,\,(j-1)!\beta_r(j)\right),
\\
{\mathcal Q}_j  & = & \frac{1}{(-1)^jj!}Y_j \left( -0!\beta_r(1),\,\,
-1!\beta_r(2)\,\,, \ldots , \,\,-(j-1)!\beta_r(j)\right).
\end{eqnarray}
As an example, let us calculate ${\mathcal P}_2$ coefficient. Using the recurrence relation (\ref{B2}) we obtain:
$
{\mathcal P}_2= (1/2)Y_2(\beta_r(1),\, \beta_r(2)) = (1/2)Y_2(\beta_r(1)^2,\, \beta_r(2))
=(1/2)\left(\prod_{j=1}^r (1-x_j^2)^{-1}+ \prod_{j=1}^r (1-x_j^2)\right).
$

{\bf The infinite hierarchy.}
Let us consider again the hierarchy $\Gamma _{-}^{\left( p\right) }(z, q)$  of
$q$-deformed vertex operators. We have
$
\Gamma _{+}\left( 1\right) \Gamma _{-}^{\left( p\right)
}\left( z, q\right) = M_{p}\left(q\right) \Gamma _{-}^{\left( p\right) }\left(
z, q\right) \Gamma _{+}\left( 1\right),
$
where $M_{p}\left(q\right)$ is precisely the generalized $p$-dimensional MacMahon function.
The general relation is the following
\begin{equation}
\left\langle 0\right\vert \Gamma _{+}\left( z_{1}\right)
\Gamma _{-}^{\left( \ell+1\right) }\left( z_{\ell}, q\right) \left\vert
0\right\rangle =\prod_{k_{\ell}=0}^\infty{\cdots }\prod_{k_{1}=0}^\infty
\prod_{k_0=0}^\infty\left[
1-q^{k_{1}+\ldots +k_{\ell}}\frac{z_{\ell}}{z_{1}}\right]^{-1},
\label{rhs}
\end{equation}
with $z_\ell/z_1 = q^{k_0}$. In the case $x_1 = x_2 = \cdots = x_\ell= q$ we get
\begin{eqnarray}
{\mathcal F}(u) & = & \prod_{{\bf k}\geq 0} \left( 1- uq^{k_1+k_2+\cdots + k_\ell}\right)^{-1}
= \exp\left( - \sum_{m=1}^\infty \frac{u^m}{m}(1-q^m)^{-\ell}\right),
\\
{\mathcal G}(u) & = & \prod_{{\bf k}\geq 0} \left( 1+ uq^{k_1+k_2+\cdots + k_\ell}\right)
= \exp\left( - \sum_{m =1}^\infty \frac{(-u)^m}{m}(1-q^m)^{-\ell}\right).
\end{eqnarray}
These formulas could be interpreted as the $\ell$ copies   of  free CFT$_{2}$ representations.
Indeed, setting $uq^{k_{1}+\ldots + k_{\ell}}=Q_{{\bf k}}q^{k_{0}}$
with $Q_{{\bf k}}=q^{k_{1}+\ldots +k_{l}}$\, (${\bf k} = \left( k_{1},\ldots ,k_{\ell}\right))$
we get
\begin{equation}
Z_{2}\left( Q_{{\bf k}},q\right) =\prod_{k_{0}=0}^{\infty}
\left[1-Q_{\bf k}q^{k_{0}}\right]^{-1} = [(1-Q_{\bf k}){\mathcal R}(s= (k_{1}+\ldots + k_{\ell})(1-i\varrho(\tau)))]^{-1}\,.
\end{equation}
Therefore the right hand side of Eq. (\ref{rhs}) can be factorized as
$
\prod_{{\bf k}\geq {\bf 0}} Z_{2}\left( Q_{{\bf k}},q\right).
$
We can treat this factorization as a product of infinite copies, each of them is $Z_{2}\left( Q_{{\bf k}},q\right)$ and corresponds to a free CFT$_2$.

\section{Conclusions}

We have shown above that all $p$-dimensional partition functions that have been considered in this paper can be written in terms of Ruelle functions, a spectral function related to hyperbolic geometry in three dimensions (see the Appendix). Thus they can not only be interpreted as correlators in a 2d CFT, but suggest a possible interpretation in terms of three dimensional physics. This relation is however still unveiled. In this last section we would like nevertheless to recall that in some specific cases the interpretation in terms three dimensional physics has been possible.

For the benefit of the reader, let us start explaining the connection between highest weight representations of infinite dimensional Lie algebras and holomorphic factorized quantum corrections for supergravity in three dimensions.
Let $M(c, h)\, (c, h \in {\mathbb C})$ be the Verma module over Virasoro (Vir) algebras. We have
$[L_0, L_{-n}] = n L_{-n}$ and $L_0$
is diagonalizable on $M(c, h)$ with spectrum $h+ {\mathbb Z}_{+}$ and with eigenspace decomposition
$
M(c, h) =\bigoplus_{j\in {\mathbb Z}_{+}} M(c, h)_{h+j}\,,
$
where $M(c, h)_{h+j}$ is spanned by elements of the basis
of $M(c, h)$. The conformal central charge $c$ acts on $M(c, h)$ as $c\,$Id.  It follows that
$
W_j = {\rm dim}\, M(c, h)_{h+j},
$
where $W_j$ is the partition function \cite{Kac}. The latter can be rewritten in the form
\begin{equation}
{\rm Tr}_{M(c, h)}\, q^{L_0} := \sum_{\lambda}q^{\lambda}{\rm dim}\,
M(c, h)_{\lambda}  = q^h\prod_{j=1}^\infty (1-q^j)^{-1}\,.
\label{ch}
\end{equation}
The series ${\rm Tr}_M\,q^{L_0}$ is called the formal character of the Vir-module $M$.

For three-dimensional gravity in a real hyperbolic space the partition function admits a factorization: it is a product of holomorphic and antiholomorphic functions
$
{W}_{0, 1}(\tau, \overline{\tau})  =
W(\tau)_{\rm hol}\cdot W(\overline{\tau})_{\rm antihol},
$
where
\begin{equation}
W(\tau)_{\rm hol} =  q^{-k}\prod_{n=1}^{\infty}(1-q^{n+1})^{-1}\,,\,\,\,\,\,\,\,\,\,
W(\overline{\tau})_{\rm antihol} =  \overline{q}^{-k}\prod_{n=1}^{\infty}(1-\overline{q}^{n+1})^{-1}\,.
\label{holomorphic}
\end{equation}
The holomorphic contribution in (\ref{holomorphic}) corresponds to the formal character of the Vir-module.

On the other hand the modulus of a Riemann surface $\Sigma$ of genus one (the conformal boundary of AdS$_3$) is defined up to $\gamma \cdot\tau = (a\tau+ b)/(c\tau + d)$ with
$\gamma \in SL(2,{\mathbb Z})$.
Therefore the generating function as the sum of known contributions
of states of left- and right-moving modes in the conformal field theory takes the form
$
\sum_{c,d}{W}_{c,d}(\tau, \overline{\tau})=
\sum_{c,d}{W}_{0,1}((a\tau+b)/(c\tau+d), \overline{\tau})\,.
$
The generating function, represented as the sum over geometries, becomes, \cite{BB},
\begin{eqnarray}
\!\!\!\!\!\!\!\!\!\!
\sum_{c,d} {W}_{c, d}(\gamma\cdot\tau, \overline{\tau}) & = &
\sum_{c,d}
\left|q^{-k}
\prod_{n=2}^{\infty}(1-q^{n})^{-1}\right|_{\gamma}^{2}
\nonumber \\
& = &
\sum_{c,d} \left\{
|q\overline{q}|^{-k}\cdot
[\cR(s=2-2i\varrho(\tau))]_{\rm hol}^{-1}\cdot
[\cR({s}=2+2i\varrho(\tau))]_{\rm antihol}^{-1}
\right\}_{\gamma}\!.
\label{summand}
\end{eqnarray}
Here $| ... |_{\gamma}$ denotes the transform of an expression $| ... |$ by $\gamma$.
The summand in (\ref{summand}) is independent of the choice of $a$ and $b$ in $\gamma$.
The sum over $c$ and $d$ in (\ref{summand}) should be thought of as a sum over the coset
$PSL(2,\mathbb Z)/{\mathbb Z} \equiv (SL(2,\mathbb Z)/\{\pm 1\})/\mathbb Z$.
This result can be extended to ${\mathcal N}= 1$ supergravity \cite{Maloney}. The infinite series of quantum
corrections for the Neveu-Schwarz and Ramond sector of supergravity can be reproduced in terms
of Ruelle spectral functions in a holomorphically factorized theory \cite{BB}.

This is an example of the fact that the Ruelle function represents a bridge between two-dimensional CFT and three dimensional physics. This might be the meaning also of the formulas we have derived in the previous sections.

In this light it is important to recall that a particular example of MacMahon function,
$Z_{3d}$, is directly linked to the topological vertex, \cite{Aganagis}, in topological string theory. This is an open topological amplitude in a Calabi-Yau background.
Analysis of this vertex, ${\mathcal C}_{\lambda\mu\nu}$,
and open string partition function leads to a relation $Z_{3d}\sim {\mathcal C}_{\lambda\mu\nu}$, \cite{Iqbal2007,Drissi2008,Drissi,Wu}. This can be achieved as
$\langle \nu^t\vert {\mathcal O}_+(\lambda){\mathcal O}_-(\lambda^t)\vert \mu\rangle$ where the operators ${\mathcal O}_+$ and ${\mathcal O}_-$ play role of composite local vertex operators of two dimensional $c=1$ conformal theory, and $\lambda, \mu, \nu$ represent boundary states described by 2d-Young diagrams. Applications of the topological vertex, see \cite{Marino}, suggest a connection with Chern-Simons theory. This may be the clue for our future investigation.

\section{Appendix: Spectral functions of hyperbolic three-geometry}
\label{Spectral}

In this section we recall some results on the Ruelle (Patterson-Selberg type) spectral functions. For details we refer the reader to \cite{BG,BCST,G} where spectral functions of hyperbolic three-geometry
were considered in connection with three-dimensional Euclidean black holes, pure supergravity, and string amplitudes.

Let ${\mathfrak G}^\gamma \in G=SL(2, {\mathbb C})$ be the discrete group defined by
\begin{eqnarray}
{\mathfrak G}^\gamma & = & \{{\rm diag}(e^{2n\pi ({\rm Im}\,\tau + i{\rm
Re}\,\tau)},\,\,  e^{-2n\pi ({\rm Im}\,\tau + i{\rm Re}\,\tau)}):
n\in {\mathbb Z}\} = \{{\gamma}^n:\, n\in {\mathbb Z}\}\,,
\nonumber \\
{\gamma} & = & {\rm diag}(e^{2\pi ({\rm Im}\,\tau + i{\rm
Re}\,\tau)},\,\,  e^{-2\pi ({\rm Im}\,\tau + i{\rm Re}\,\tau)})\,.
\label{group}
\end{eqnarray}
One can construct a zeta function of Selberg-type for the group
${\mathfrak G}^\gamma \equiv {\mathfrak G}_{(\alpha, \beta)}^\gamma$ generated by a single hyperbolic
element of the form ${\gamma_{(\alpha, \beta)}} = {\rm diag}(e^z, e^{-z})$,
where $z= \alpha +i\beta$ for $\alpha, \beta >0$. Actually $\alpha = 2\pi {\rm Im}\,\tau$ and $\beta = 2\pi {\rm Re}\,\tau$.
The Patterson-Selberg spectral function $Z_{{\mathfrak G}^\gamma} (s)$ and its logarithm for ${\rm Re}\, s> 0$ can be attached
to $H^3/{\mathfrak G}^\gamma$ as follows:
\begin{eqnarray}
Z_{{\mathfrak G}^\gamma}(s) & := & \prod_{k_1,k_2\geq
0}[1-(e^{i\beta})^{k_1}(e^{-i\beta})^{k_2}e^{-(k_1+k_2+s)\alpha}]\,,
\label{zeta00}
\\
{\rm log}\, Z_{{\mathfrak G}^\gamma} (s) \!\! & = &
-\frac{1}{4}\sum_{n = 1}^{\infty}\frac{e^{-n\alpha(s-1)}}
{n[\sinh^2\left(\frac{\alpha n}{2}\right)
+\sin^2\left(\frac{\beta n}{2}\right)]}\,.
\label{logZ}
\end{eqnarray}
For more information about the analytic properties
of this spectral function we refer the reader to the papers \cite{Patterson,BCST}.
Let us introduce the Ruelle functions $\cR(s)$, as an alternating product of factors, each
of which is a Selberg zeta function ($\cR(s)$ and $Z_{{\mathfrak G}^\gamma}(s)$ can be continued meromorphically to the entire
complex plane $\mathbb C$.)
\begin{eqnarray}
\prod_{n=\ell}^{\infty}(1- q^{an+\varepsilon})
& = & \prod_{p=0, 1}Z_{{\mathfrak G}^\gamma}(\underbrace{(a\ell+\varepsilon)(1-i\varrho(\tau))
+ 1 -a}_s + a(1 + i\varrho(\tau)p)^{(-1)^p}
\nonumber \\
& = &
\cR(s = (a\ell + \varepsilon)(1-i\varrho(\tau)) + 1-a),
\label{R1}
\\
\prod_{n=\ell}^{\infty}(1+ q^{an+\varepsilon})
& = &
\prod_{p=0, 1}Z_{{\mathfrak G}^\gamma}(\underbrace{(a\ell+\varepsilon)(1-i\varrho(\tau)) + 1-a +
i\sigma(\tau)}_s
+ a(1+ i\varrho(\tau)p)^{(-1)^p}
\nonumber \\
& = &
\cR(s = (a\ell + \varepsilon)(1-i\varrho(\tau)) + 1-a + i\sigma(\tau))\,,
\label{R2}
\end{eqnarray}
being $q\equiv e^{2\pi i\tau}$, $\varrho(\tau) = {\rm Re}\,\tau/{\rm Im}\,\tau= \beta/\alpha$,
$\sigma(\tau)= 1/(2\,{\rm Im}\,\tau) = \pi/\alpha$, while $a$ is a real number, $\ell \in {\mathbb Z}_+$  and $\varepsilon \in {\mathbb C}$.
We can use the Ruelle functions $\cR(s)$ to write the results in a most general form
\begin{eqnarray}
\!\!\!\!\!\!
\prod_{n=\ell}^{\infty}(1- q^{an+\varepsilon})^{C(n)} & = & \prod_{n=\ell}^{\infty}\left[
\frac{{\mathcal R}(s= (an+\epsilon)(1-i\varrho(\tau))+1-a)}
{{\mathcal R}(s= (a(n+1)+\epsilon)(1-i\varrho(\tau))+1-a}\right]^{C(n)}\!\!\!\!\!\!\!\!\!,
\label{G1}
\\
\!\!\!\!\!\!
\prod_{n=\ell}^{\infty}(1+ q^{an+\varepsilon})^{C(n)} & = & \prod_{n=\ell}^{\infty}\left[
\frac{{\mathcal R}(s= (an+\epsilon)(1-i\varrho(\tau))+1-a + i\sigma(\tau))}
{{\mathcal R}(s= (a(n+1)+\epsilon)(1-i\varrho(\tau))+1-a + i\sigma(\tau)}\right]^{C(n)}\!\!\!\!\!\!\!\!\!,
\label{G2}
\end{eqnarray}
where $C(n)$ are certain coefficients. In the simplest cases $C(n)=bn$, $b\in {\mathbb R}$,
and eqs. (\ref{G1}) and (\ref{G2}) becomes
\begin{eqnarray}
\prod_{n=\ell}^{\infty}(1-q^{an+ \epsilon})^{bn} & = &
\cR(s=(a\ell + \varepsilon)(1-i\varrho(\tau))+1-a)^{b\ell}
\nonumber \\
& \times &
\!\!\!
\prod_{n=\ell+1}^{\infty}
\cR(s=(an + \varepsilon)(1-i\varrho(\tau))+1-a)^{b}\,,
\label{R3}
\\
\prod_{n=\ell}^{\infty}(1+q^{an+ \epsilon})^{bn} & = &
\cR(s=(a\ell + \varepsilon)(1-i\varrho(\tau))+1-a+ i\sigma(\tau))^{b\ell}
\nonumber \\
& \times &
\!\!\!
\prod_{n=\ell+1}^{\infty}
\cR(s=(an + \varepsilon)(1-i\varrho(\tau))+1-a+ i\sigma(\tau))^{b}\,.
\label{R4}
\end{eqnarray}

\subsection*{Acknowledgements}
AAB would like thank ICTP and SISSA (Italy), and CNPq (Brazil) for financial support and SISSA for hospitality during this research.


\begin{thebibliography}{99}


\bibitem{Okunkov}
A. Okounkov, N. Reshetikhin and C. Vafa, {\it Quantum Calabi-Yau and Classical Crystals},
Progr. Math. {\bf 244} (2006) 597-618; [arXiv:hep-th/0309208].

\bibitem{Kenyon2003}
R. Kenyon, A. Okounkov and S. Sheffield, {\it Dimers and Amoebae}, arXiv:math-ph/0311005.

\bibitem{Kenyon2006}
R. W. Kenyon and D. B. Wilson, {\it Boundary Partitions in Trees and Dimers}, Trans. Am. Math. Soc.
{\bf 363} (2011) 1325-1364; [arXiv:math.CO/0608422].

\bibitem{Ghostal}
D. Ghoshal and C. Vafa, {\it c =1 string as the topological
theory of the conifold}, Nucl. Phys. B {\bf 453} (1995) 121-128;
[arXiv:hep-th/9506122].

\bibitem{Witten}
E. Witten, {\it Ground ring of two-dimensional string theory}, Nucl. Phys. B {\bf 373} (1992) 187-213;
[arXiv:hep-th/9108004].

\bibitem{Saidi}
E. H. Saidi and  M. B. Sedra, {\it Topological string in harmonic space and correlation functions
in $S^3$ stringy cosmology}, Nucl. Phys. B {\bf 748} (2006) 380-457; [arXiv:hep-th/0604204].

\bibitem{Douglas}
M. R. Douglas and G. Moore, {\it D-branes, Quivers, and ALE Instantons}, arXiv:hep-th/9603167v1.

\bibitem{Benhaddou}
M. A. Benhaddou and  E. H. Saidi, {\it Explicit Analysis of Kahler Deformations in 4D N=1 Supersymmetric
Quiver Theories}, Phys. Lett. B {\bf 575} (2003) 100-110; [arXiv:hep-th/0307103].

\bibitem{Saraikin}
K. Saraikin and C. Vafa, {\it Non-supersymmetric Black Holes and Topological Strings},
Class. Quant. Grav. {\bf 25} (2008) 095007; [arXiv:hep-th/0703214].

\bibitem{Aganagis}
M. Aganagic, A. Klemm, M. Marino and C. Vafa, {\it The Topological Vertex},
Commun. Math. Phys. {\bf 254} (2005) 425-478; [arXiv:hep-th/0305132].

\bibitem{Iqbal2003}
A. Iqbal, N. Nekrasov, A. Okounkov and C. Vafa, {\it Quantum Foam and Topological Strings},
JHEP {\bf 0804} (2008) 011; [arXiv:hep-th/0312022].

\bibitem{Iqbal2007}
A. Iqbal, C. Kozcaz and C. Vafa, {\it The Refined Topological Vertex},
JHEP {\bf 0910} (2009) 069; [arXiv:hep-th/0701156].

\bibitem{Drissi}
L. B. Drissi, J. Houda and E. H. Saidi, {\it Refining the Shifted Topological Vertex},
J. Math. Phys. {\bf 50} (2009) 013509; [arXiv:hep-th/0812.0513].

\bibitem{BB}
L. Bonora and A. A. Bytsenko, {\it Partition functions for quantum gravity, black holes,
elliptic genera and Lie algebra homologies}, Nucl. Phys. B {\bf 852} (2011) 508–537;
[arXiv:hep-th/1105.4571].

\bibitem{BBE}
L. Bonora, A. A. Bytsenko and E. Elizalde, {\it String partition functions, Hilbert schemes
and affine Lie algebra representations on homology groups}, J. Phys. A {\bf 45} (2012)
374002; [arXiv:hep-th/1206.0664].

\bibitem{Kac}
V. G. Kac, {\it Infinite Dimensional Lie Algebras}, third ed. Cambridge University Press, 1990.

\bibitem{Awata}
H. Awata, M. Fukuma, Y. Matsuo, and S. Odake, {\it Representation Theory of the $W_{1+\infty }$ Algebra},
Prog. Theor. Phys. Suppl. {\bf 118} (1995) 343-374; [arXiv:hep-th/9408158].

\bibitem{Frenkel}
E. Frenkel, V. Kac, A. Radul and W.-Q. Wang, {\it $W_{1+\infty }$ and $W(gl_{N})$ with Central Charge N},
Comm. Math. Phys. {\bf 170} (1995) 337-358; [arXiv:hep-th/9405121].

\bibitem{Heckman}
J. J. Heckman and  C. Vafa, {\it Crystal Melting and Black Holes}, JHEP {\bf 0709} (2007) 011;
[arXiv:hep-th/0610005].

\bibitem{Drissi2008}
L. B. Drissi, J. Houda and E. H. Saidi, {\it Generalized MacMahon $G_d(q)$ as $q$-deformed CFT$_2$
correlation function}, Nucl. Phys. B {\bf 801} (2008) 316-345; [arXiv:hep-th/0801.2661v2].

\bibitem{Andrews}
G. E. Andrews, {\it The Theory of Partitions}, Encyclopedia of Mathematics vol. 2, Addison-Wesley
Publishing Company, 1976.

\bibitem{Maloney}
A. Maloney and E. Witten, {\it Quantum Gravity Partition Function In Three Dimensions},
JHEP {\bf 1002} (2010) 029; [arXiv:hep-th/0712.0155].

\bibitem{Wu}
J.-f. Wu and J. Yang, {\it Vertex Operators, ${\mathbb C}^3$ Curve, and Topological Vertex},
arXiv:hep-th/1403.0181v1.

\bibitem{Marino} M.~Marino, {\it Chern-Simons theory and topological strings,} Rev.\ Mod.\ Phys.\  {\bf 77} (2005) 675-720;
[arXiv:hep-th/0406005].

\bibitem{BG}
A. A. Bytsenko and M. E. X. Guimar\~aes, {\it Truncated Heat Kernel and One-Loop Determinants for the BTZ Geometry},
Eur. Phys. J.  C {\bf 58} (2008) 511-516; [arXiv:hep-th/0809.1416].

\bibitem{BCST}
A. A. Bytsenko, M. Chaichian, R. J. Szabo and A. Tureanu, {\it Quantum Black Holes, Elliptic Genera and Spectral Partition Functions},
to appear in the IJGMMP (2014); [arXiv:hep-th/1308.2177].

\bibitem{G}
M. E. X. Guimar\~aes, R. M. Luna and T. O. Rosa, {\it Topological Vertex, String Amplitudes and Spectral Functions of Hyperbolic Geometry},
to appear in the Eur. Phys. J.  C (2014); [arXiv:hep-th/1403.7139].

\bibitem{Patterson}
S. J. Patterson and P. A. Perry, {\it The divisor of the Selberg zeta function for Kleinian groups, with an appendix by Charles Epstein}, Duke Math. J. {\bf 106} (2001) 321-390.


\end{thebibliography}
\end{document}